\documentclass[preprint,3p,times]{elsarticle}

\usepackage{amssymb,amsmath,amsfonts,mathrsfs,bm,gensymb}
\usepackage{graphicx,float,subfig}
\usepackage{lineno}

\usepackage{geometry}
\usepackage{hyperref}
\usepackage{tabularx}
\usepackage{array}
\usepackage{color}

\newcolumntype{P}[1]{>{\centering\arraybackslash}p{#1}}
\newcommand\ChangeRT[1]{\noalign{\hrule height #1}}

\definecolor{gray}{RGB}{170, 170, 170}

\hypersetup{
    colorlinks=true,
    linkcolor=black,
    citecolor=black,
    filecolor=black,
    urlcolor=black,
}

\makeatother

\biboptions{sort&compress}

\begin{document}

\begin{frontmatter}

\title{
Time-modulated inerters as building blocks for nonreciprocal mechanical devices
}

\author[add2]{P. Celli\corref{corr1}}
\ead{paolo.celli@stonybrook.edu}
\author[add1]{A. Palermo\corref{corr1}}
\ead{antonio.palermo6@unibo.it}
\cortext[corr1]{Equal contributors and corresponding authors}

\address[add2]{Department of Civil Engineering, Stony Brook University, Stony Brook, NY 11794, USA}
\address[add1]{Department of Civil, Chemical, Environmental and Materials Engineering, University of Bologna, 40136 Bologna, Italy}

\begin{abstract}

In this work, we discuss the realization of mechanical devices with non-reciprocal attributes enabled by inertia-amplifying, time-modulated mechanisms. Our fundamental building-block features a mass, connected to a fixed ground through a spring and to a moving base through a mechanism-based inerter. Through analytical derivations and numerical simulations, we provide details on the nonlinear dynamics of such system. We demonstrate that providing a time modulation to the inerter's base produces two additions on the dynamics of the main spring-mass oscillator: i) an effective time-modulated mass term, and ii) a time varying force term; both quantities are functions of the modulating frequency. With specific choices of parameters, the modulation-induced force term -- that represents one of the main drawbacks in most experimental realizations of purely time-modulated systems -- vanishes and we are left with an effective time-varying mass. We then illustrate that this building block can be leveraged to realize non-reciprocal wave manipulation devices, and concentrate on a non-reciprocal beam-like waveguide. The simple design and the clean performance of our system makes it an attractive candidate for the realization of fully mechanical non-reciprocal devices.

\vspace{5px}
\normalsize{\textbf{This article may be downloaded for personal use only. Any other use requires prior permission of the author and Elsevier. This article appeared in}: \emph{Journal of Sound and Vibration} 572, 118178 (2024) \textbf{and may be found at}: \url{https://doi.org/10.1016/j.jsv.2023.118178}}
\end{abstract}


\end{frontmatter}

\section{Introduction}
\label{s:intro}


Mechanical systems and structures with time-varying properties have been of interest to the mechanics and nonlinear dynamics communities for decades. Some of these property variations are unwanted, as it is the case in rotordynamics~\cite{childs1979rub}. Since the roller bearings supporting a shaft provide a harmonically varying stiffness, parametric excitations can arise and lead to instabilities~\cite{Tehrani2020}. In other cases, such property variations produce desired effects, as in the inverted pendulum with a vibrating base~\cite{kapitza1965dynamical}. In this system, the typically-unstable inverted configuration is stabilized by vibrations of its base. Recently, time-varying properties have also been leveraged in phononic crystals and metamaterials to manipulate mechanical waves in exotic ways~\cite{nassar2020}. In phononic crystals, characterized by spatially-periodic material properties, synchronous time-modulations have yielded wavenumber badgaps~\cite{moghaddaszadeh2022complex}. Spatio-temporal modulations, on the other hand, have yielded non-reciprocal effects~\cite{Swinteck2015, Trainiti2016}, which are particularly sought-after to realize unidirectional transmission devices~\cite{goldsberry2022nonreciprocity}. The same applies to metamaterials, where the time-varying characteristics are provided to the  resonators attached to a wave-carrying medium~\cite{Nassar2017prsa, Nassar2017eml}. In the context of metamaterials, such effects stem from the dynamics of single mechanical resonators with time-varying properties; just like their optical counterparts~\cite{Minkov2017}, modulated mechanical resonators display sidebands, peaks in the frequency response that are shifted at integer multiples of the modulated frequency from the main resonance peak~\cite{Palermo2020}. All these systems with modulated properties can be modeled using equations of the Mathieu type~\cite{Ruby1996, Kovacic2018}.


Here, we are interested in time-modulations and on how to achieve them in fully-mechanical systems. So far, property modulations have been mainly introduced via smart materials and structures. Early realizations of non-reciprocal, time-modulated media for mechanical wave control featured electromagnets~\cite{Wang2018, Chen2019}. Much work has also been done with shunted piezoelectrics~\cite{trainiti2019, Marconi2020, sugino2020nonreciprocal}. Additionally, Attarzadeh et al.\ have leveraged miniature electric motors to modulate the apparent moment of area of cantilever-beam resonators attached on a beam waveguide~\cite{Attarzadeh2020}. While these systems indeed display non-reciprocal attributes, the quest for fully-mechanical time-modulated systems is still ongoing. Goldsberry et al.\ proposed a continuous elastic waveguide with non-reciprocal attributes stemming from negative stiffness effects~\cite{Goldsberry2019}, while Wang et al.\ discussed non-reciprocity in a tensegrity chain with modulated prestress~\cite{wang2020prestress}  \textcolor{black}{and Rosi{\'c} et al.\ proposed a system of laterally-forced beams connected via viscoelastic layers~\cite{rosic2022parametrically}}. 
A challenge associated with many of these structures is that introducing property modulations also introduce forces that generate large-amplitude modulation waves in the system; in other words, the amplitude of the signal to be manipulated is a fraction of that of the modulated wave. As a side note, and to complete our literature review, researchers have also explored a parallel avenue to achieve non-reciprocity by leveraging nonlinearities rather than time-modulations~\cite{Boechler2011, lu2020non, yu2022non}.


Recent works have also explored the use of devices with mechanism-like behaviors to produce effective modulations in stiffness and mass~\cite{huang2020non, karlivcic2022non}. Here, a mechanism is intended as a zero-energy mode of deformation~\cite{Pellegrino1986}; while these are known as collapse mechanisms and yield instability in structures, they are currently leveraged in robotics and other applications to create linkages that can follow desired motion paths at very low energetic costs. Huang et al.\ introduced a spring-mass system where each mass is connected to two spinning structures; one of them features auxiliary masses and acts as an inerter, thereby causing the main mass to effectively be time-modulated, while the other introduces a time-modulated stiffness~\cite{huang2020non}. Karli\v{c}i\'{c} et al.\ recently studied a similar system, where 2D mechanism-based inerters are arranged in a 1D chain and modeled as time-varying inertances~\cite{karlivcic2022non}. Inerters can in fact be realized by means of four-bar linkages with masses at the nodes~\cite{Yilmaz2007}, or by compliant analogs~\cite{Acar2013, Taniker2015}. When one node of the linkage is actuated, the motion of the masses located at the adjacent nodes produces an inertial force on the opposite node of the linkage. In turn, this produces an angle-dependent mass term in the equation of motion of the inerter which can amplify the effective mass of the device. Indeed, the linear and non-linear dynamics of discrete chains and waveguides equipped with such inertia-amplifying mechanisms have been the object of multiple studies ~\cite{Yilmaz2010,Zeighami2019,Settimi2021,Settimi2023}.
Nonetheless, the same inerters are promising in the context of time-modulated systems since they allow to easily produce time-varying masses. In our work, we highlight  a few more reasons why such devices deserve attention, which will become clear by carefully analyzing its nonlinear dynamics.  

\begin{figure}[!htb]
\centering
\includegraphics[scale=1.00]{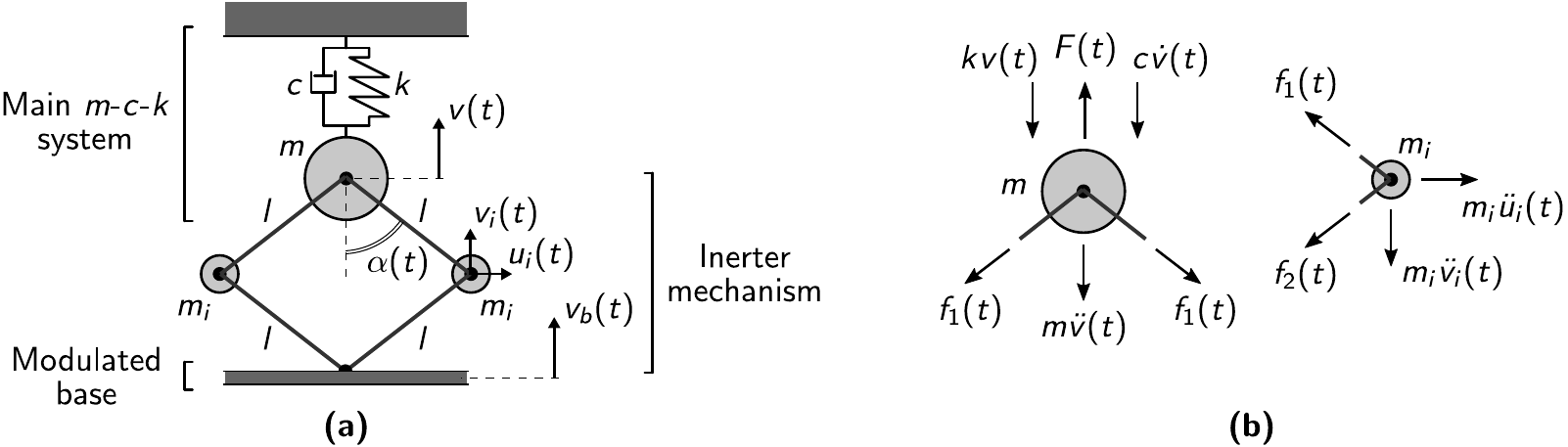}
\caption{(a) Schematic of our building block, with all its characteristic quantities. It features a main spring-mass-damper system, connected to a movable base through an inertial amplification mechanism. The black lines are inextensible bars, the black circles are perfect pin connectors and the large gray circles are point masses. (b) Free-body diagram for the main mass (left) and a lateral mass (right), indicating bar tensions $f_1$ and $f_2$.}
\label{f:block}
\end{figure}

In what follows we provide a detailed study on the dynamics of a time-modulated building block composed of a single degree of freedom (SDOF) system featuring a main mass, connected to a fixed base through a spring and to a modulated base through an inerter. A schematic of such system is illustrated in Fig.~\ref{f:block}(a).
The fact that the inerter is connected to a moving base is what makes our system unique. By investigating the nonlinear equation of motion of this system, we identify two key contributions introduced by the base modulation: an effective time-varying mass and a dynamic force term. With an eye towards possible experimental realizations of such system, we provide detailed information on the parameters driving its response. In particular we discuss the parametric range to be chosen so that the dynamic force term disappears and the system can be simply analyzed by considering its time-varying mass. Then, we illustrate that this building block can be used to create a beam waveguide with non-reciprocal attributes. Throughout our work, analytical findings are compared to fully-nonlinear finite element simulations that capture the large configuration change in our mechanism-based devices.


The article is organized as follows. In Section~\ref{s:bb}, we introduce our building block, study its free and forced response, and discuss the influence of various design parameters on its response and performance. In Section~\ref{s:ass}, we investigate how arrays of our building blocks can be used to create a beam waveguide with non-reciprocal attributes. The conclusions of our work are laid out in Section~\ref{s:concl}.

\section{The time-modulated building block}
\label{s:bb}

In this section, we analyze the dynamics of the spring-mass system equipped with a base-modulated inerter by means of analytical derivations and numerical simulations. 

\subsection{Properties of the system and nonlinear equation of motion}
\label{s:eom}

Our building block, with all the relevant quantities of interest, is schematically illustrated in Fig.~\ref{f:block}(a). The system features a main mass $m$, connected to ground through a spring of stiffness $k$ and a damper with damping coefficient $c$. The mass can move only along the vertical direction with displacement $v(t)$ and is subjected to a harmonic force $F(t)$, where $t$ indicates time. The same mass is connected to a movable base through a displacement-amplifying mechanism, that features four rigid bars of length $l$ connected by four frictionless pin joints. The top pin overlaps with the main mass, $m$. The bottom one is connected to the moving base, that oscillates vertically by a displacement $v_b(t)$. Each lateral pin features a point mass of magnitude $m_i$, whose vertical and horizontal displacements are labeled $v_i(t)$ and $u_i(t)$, respectively. These displacements are identical in magnitude between the two masses, due to the symmetries of our structure. Each bar stemming from $m$ is tilted with respect to the vertical direction by an angle $\alpha(t)$; when the structure is undeformed, this angle is called $\alpha_0$. In the following, to be concise, we will omit most time dependencies: $F(t)=F$, $v(t)=v$, $v_b(t)=v_b$, $\alpha(t)=\alpha$, $u_i(t)=u_i$ and $v_i(t)=v_i$. 

By analyzing the kinematics of such building block, we can define the following expressions:
\begin{equation}
    2l\cos{\alpha}=2l\cos{\alpha_0}+v-v_b,\quad u_i=l\left(\sin{\alpha_0}-\sin{\alpha}\right),\quad v_i=v_b+l\left(\cos{\alpha}-\cos{\alpha_0}\right).
    \label{e:k}
\end{equation}
From the equilibrium of forces about one of the lateral masses, we obtain the tensions along the bars above and below the lateral masses ($f_1$ and $f_2$, respectively, as shown in Fig.~\ref{f:block}(b)):
\begin{equation}
    f_1=m_i\left( \frac{\ddot{u}_i}{2\sin{\alpha}} + \frac{\ddot{v}_i}{2\cos{\alpha}} \right),\qquad f_2=m_i\left( \frac{\ddot{u}_i}{2\sin{\alpha}} - \frac{\ddot{v}_i}{2\cos{\alpha}} \right),
    \label{e:f}
\end{equation}
where the overdot indicates a time derivative. Note that these forces vary as a function of time.

From the equilibrium of vertical forces acting on the main mass (the horizontal forces are in equilibrium provided that the two lateral masses and all the lengths of the arms of the mechanism are identical), and using Eq.~\ref{e:f}, we can write the following equation of motion:
\begin{equation}
    m_i\frac{{\ddot{u}_i}\,\cos{\alpha}+{\ddot{v}_i}\,\sin{\alpha}}{\sin{\alpha}}+m\ddot{v}+c\dot{v}+kv=F.
\end{equation}
Using the definitions of $\alpha$, $u_i$ and $v_i$ obtained through kinematic analysis (Eq.~\ref{e:k}), we can spell out each term in the equation as follows:
\begin{multline}
    \underbrace{\left(m+\frac{m_i}{2}\right){\ddot{v}}}_{\mathrm{T}_1} + 
    \underbrace{\frac{m_i}{2}{\ddot{v}_b}}_{\mathrm{T}_2} +
    \underbrace{\frac{m_i}{2}\frac{\left( \dot{v} - \dot{v}_b \right)^2\left( v-v_b+2\,l\cos{\alpha_{0}} \right)}{4\,l^2-{\left(v-v_b+2l\cos{\alpha_0}\right)}^2}}_{\mathrm{T}_3} \\ +
    \underbrace{\frac{m_i}{2}\frac{\left( \dot{v} - \dot{v}_b \right)^2\left( v-v_b+2\,l\cos{\alpha_{0}} \right)^3}{\left(4\,l^2-{\left(v-v_b+2\,l\cos{\alpha_0}\right)}^2 \right)^2}}_{\mathrm{T}_4} +
    \underbrace{\frac{m_i}{2}\frac{\left( \ddot{v} - \ddot{v}_b \right)\left( v-v_b+2\,l\cos{\alpha_{0}} \right)^2}{4\,l^2-{\left(v-v_b+2\,l\cos{\alpha_0}\right)}^2}}_{\mathrm{T}_5} + \underbrace{c\dot{v}}_{\mathrm{T}_6} + \underbrace{kv}_{\mathrm{T}_7}=F,
    \label{e:nleom}
\end{multline}
where we labeled all terms of the equation from $\mathrm{T}_1$ to $\mathrm{T}_7$, for future reference.

Assuming a sinusoidal base excitation at $\omega_m$, the modulating frequency, we can define:
\begin{equation}
    v_b(t)=dv_b\,\cos(\omega_m\,t),\quad\dot{v}_b(t)=-dv_b\,\omega_m\,\sin(\omega_m\,t),\quad\ddot{v}_b(t)=-dv_b\,\omega_m^2\,\cos(\omega_m\,t),
\end{equation}
where $dv_b$ is the amplitude of base motion. Substituting into Eq.~\ref{e:nleom}, we obtain a nonlinear equation that can be solved numerically:
\begin{multline}
    \left(m+\frac{m_i}{2}\right){\ddot{v}}  - 
    \frac{m_i}{2}dv_b\,\omega_m^2\,\cos(\omega_m\,t) +
    \frac{m_i}{2}\frac{\left( \dot{v} + dv_b\,\omega_m\,\sin(\omega_m\,t) \right)^2\left( v-dv_b\,\cos(\omega_m\,t)+2\,l\cos{\alpha_{0}} \right)}{4\,l^2-{\left(v-dv_b\,\cos(\omega_m\,t)+2l\cos{\alpha_0}\right)}^2} \\ +
    \frac{m_i}{2}\frac{\left( \dot{v} + dv_b\,\omega_m\,\sin(\omega_m\,t) \right)^2\left( v-dv_b\,\cos(\omega_m\,t)+2\,l\cos{\alpha_{0}} \right)^3}{\left(4\,l^2-{\left(v-dv_b\,\cos(\omega_m\,t)+2\,l\cos{\alpha_0}\right)}^2 \right)^2} \\ +
    \frac{m_i}{2}\frac{\left( \ddot{v} + dv_b\,\omega_m^2\,\cos(\omega_m\,t) \right)\left( v-dv_b\,\cos(\omega_m\,t)+2\,l\cos{\alpha_{0}} \right)^2}{4\,l^2-{\left(v-dv_b\,\cos(\omega_m\,t)+2\,l\cos{\alpha_0}\right)}^2} + c\dot{v} + kv=F.
    \label{e:nleom2}
\end{multline}
We set this equation aside and will use it later in this section to provide a comparison to our analytical results.

\subsection{Simplification of the equation of motion}
\label{s:simp}

In the following, we introduce assumptions on the nature of the base excitation and on the system at hand to try and simplify the nonlinear equation of motion in Eq.~\ref{e:nleom}. In particular, our goals are to i) derive an expression where the contribution of the time-modulated inerter is condensed into an effective time-varying mass, and ii) understand which conditions make this assumption acceptable.

First, we assume that $v \ll v_b$. Note that this is different from a classical small-on-large scenario, where the output $v$ comprises both the large modulating and the small modulated signals. In our case, conversely, the output signal is small in amplitude with respect to the modulating input $v_b$. This is made possible by the fact that the inerter, being a mechanism, does not directly transfer the base motion to the resonator. 
To enforce this assumption, we need to make sure that the initial conditions are small in a free response problem and that the force applied to the mass is small in a forced problem. In addition, we assume that we are operating the system at a frequency (the natural frequency in a free response or the excitation frequency in a forced scenario) that is larger than the modulation frequency $\omega_m$. Under these assumptions, we can deduce that $|(\dot{v}-\dot{v}_b)^2| \ll |\ddot{v}-\ddot{v}_b|$ and that terms $\mathrm{T}_3$ and $\mathrm{T}_4$ in Eq.~\ref{e:nleom} can be dropped. This simplification is further explained in~\ref{a:simp}. After neglecting $v$ in $\mathrm{T}_5$, we are left with
\begin{equation}
    \left(m+\frac{m_i}{2}\right){\ddot{v}} + 
    \frac{m_i}{2}{\ddot{v}_b} +
    \frac{m_i}{2}\frac{\left( \ddot{v} - \ddot{v}_b \right)\left( 2\,l\cos{\alpha_{0}}-v_b\right)^2}{4\,l^2-{\left(2\,l\cos{\alpha_0}-v_b\right)}^2} + c\dot{v}+kv=F.
    \label{e:seom0}
\end{equation}
This equation can be rearranged to identify an effective mass and an additional effective force. We obtain the following expression:
\begin{equation}
    \underbrace{\left( m+m_i\frac{2\,l^2}{4\,l^2-{\left(2\,l\cos{\alpha_0}-v_b\right)}^2} \right)}_{m_e(t)} \ddot{v} + c\dot{v} +kv = \underbrace{-m_i \frac{2\,l^2 -\left( 2\,l\cos{\alpha_{0}}-v_b \right)^2}{4\,l^2-{\left(2\,l\cos{\alpha_0}-v_b\right)}^2} \ddot{v}_b}_{F_e(t)} +F.
    \label{e:seom}
\end{equation}
From this expression, we can draw a significant conclusion. A time-modulated inerter has two main effects on the equation of motion of a mass that is connected to it: i) it causes the system to display a time-varying effective mass $m_e=m_e(t)$; and ii) it produces an effective force $F_e=F_e(t)$ on the mass. Assuming sinusoidal base motion, i.e., $v_b(t)=dv_b\,\cos(\omega_m\,t)$, Eq.~\ref{e:seom} can be rewritten in a form that can be solved numerically:
\begin{multline}
    \underbrace{\left( m+m_i\frac{2\,l^2}{4\,l^2-{\left(2\,l\cos{\alpha_0}-dv_b\,\cos(\omega_m\,t)\right)}^2} \right)}_{m_e(t)} \ddot{v} + c\dot{v} +kv\\ =F + \underbrace{m_i\, \frac{2\,l^2 -\left( 2\,l\cos{\alpha_{0}}-dv_b\,\cos(\omega_m\,t) \right)^2}{4\,l^2-{\left(2\,l\cos{\alpha_0}-dv_b\,\cos(\omega_m\,t)\right)}^2}\, dv_b\,\omega_m^2\,\cos(\omega_m\,t)}_{F_e(t)}.
    \label{e:seom2}
\end{multline}

It is now interesting to compare our simplified equation to the known linearized equation of inerter-based mechanical systems~\cite{Yilmaz2010}. Substituting the first kinematic relationship of Eq.~\ref{e:k} (neglecting $v\ll v_b$) in the last term of Eq.~\ref{e:seom0}, and through algebraic manipulation, we can write
\begin{equation}
    \left(m+\frac{m_i}{2}\right){\ddot{v}} + 
    \frac{m_i}{2}{\ddot{v}_b} +
    \frac{m_i}{2}\frac{\left( \ddot{v} - \ddot{v}_b \right)}{\tan^2{\alpha}} + c\dot{v}+kv=F,
    \label{e:yeom}
\end{equation}
where $\alpha=\alpha(t)$. This equation is almost identical to the one proposed by Yilmaz and Hulbert for inerter-based phononic structures~\cite{Yilmaz2010}; the only difference is the presence of  $\alpha(t)$ instead of $\alpha_0$, that stems from the time-modulated nature of our inerter.

\subsection{Analytical solution and numerical validation}
\label{s:analy}

To derive an analytical solution of the linearized version of our equation of motion (Eq.~\ref{e:seom} or Eq.~\ref{e:seom2}), we follow a procedure akin to the one typically used when analyzing the response of mechanical systems with time-varying coefficients~\cite{Vila2017, Palermo2020}. The presence of a modulating force $F_e(t)$ -- germane to our problem of inerter-based modulation -- presents some additional challenges with respect to previous literature. 

We begin by re-writing Eq.~\ref{e:seom2} in the following compact form with spelled-out time dependencies:
\begin{equation}
    m_e(t)\ddot{v}(t)+c\dot{v}(t)+kv(t)=F(t) + F_e(t).
    \label{e:seoms}
\end{equation}
Since $m_e(t)$ and $F_e(t)$ are time-periodic, we can write them in Fourier series form as
\begin{equation}
m_e(t)=\sum_{p=-\infty}^{\infty}\hat{m}_{p}\,e^{i p \omega_m t}
\label{e:me}
\end{equation}
and
\begin{equation}
F_e(t)=\sum_{p=-\infty}^{\infty}\hat{F}_{ep}\,e^{i p \omega_m t},
\label{e:fe}
\end{equation}
with Fourier coefficients defined as
\begin{equation}
\hat{m}_{p}=\frac{\omega_m}{2\pi}\int_{-\frac{\pi}{\omega_m}}^{\frac{\pi}{\omega_m}} \left( m+m_i\frac{2\,l^2}{4\,l^2-{\left(2\,l\cos{\alpha_0}-dv_b\,\cos(\omega_m\,t)\right)}^2} \right)\,e^{-ip\omega_mt} dt
\label{e:mecoef}
\end{equation}
and
\begin{equation}
    \hat{F}_{ep}=\frac{\omega_m}{2\pi}\int_{-\frac{\pi}{\omega_m}}^{\frac{\pi}{\omega_m}} m_i\, \frac{2\,l^2 -\left( 2\,l\cos{\alpha_{0}}-dv_b\,\cos(\omega_m\,t) \right)^2}{4\,l^2-{\left(2\,l\cos{\alpha_0}-dv_b\,\cos(\omega_m\,t)\right)}^2}\, dv_b\,\omega_m^2\,\cos(\omega_m\,t)\,e^{-ip\omega_mt} dt.
\label{e:fecoef}
\end{equation}

As a first step in our analysis, we set all forcing terms to zero, a scenario that is akin to classical free vibrations. We then restore the forcing terms and derive a full solution to the forced problem.

\subsubsection{Response without forcing terms: Analytical derivations}
\label{s:free}

By setting all forcing terms to zero, the equation of motion can be re-written as follows:
\begin{equation}
    m_e(t)\ddot{v}(t)+c\dot{v}(t)+kv(t)=0.
    \label{e:seomsf}
\end{equation}
Note that the solution to this equation is not a true free response of the system since: i) it accounts for the harmonic motion of the base of the inerter via $m_e(t)$; and ii) $F_e(t)$, which also stems from the same base motion, is not considered in this equation. In spite of this, solving Eq.~\ref{e:seomsf} will allow us to derive the eigensolution of the system.

Eq.~\ref{e:seomsf} can be solved by assuming a harmonic solution with time modulated amplitude, of the form
\begin{equation}
v(t)=\left(\sum_{n=-\infty}^{\infty}\hat{v}_{n}\,e^{i n \omega_m t}\right)e^{i \omega t},
\label{e:v}
\end{equation}
with $\hat{v}_{n}$ being the Fourier coefficients of the displacement function and $\omega$ being a generic angular frequency.

Differentiating $v(t)$, plugging it into Eq.~\ref{e:seomsf} together with Eq.~\ref{e:me} and simplifying $e^{i\omega t}$, yields
\begin{equation}
    -\sum_{n=-\infty}^{\infty}\sum_{p=-\infty}^{\infty}\left( \omega+n\omega_m \right)^2\hat{m}_{p}\hat{v}_{n}\,e^{i (n+p) \omega_m t}+\sum_{n=-\infty}^{\infty}\left(k+ic\left( \omega+n\omega_m \right)\right)\,\hat{v}_{n}\,e^{i n \omega_m t}=0.
    \label{e:series}
\end{equation}
As done in previous works ~\cite{Trainiti2016, Vila2017, Attarzadeh2018, Palermo2020}, we pre-multiply this expression by $e^{-iq\omega_m t}\omega_m/(2\pi)$, where $q$ is an arbitrary integer, and we integrate the result over the modulation period, from $-\pi/\omega_m$ to $\pi/\omega_m$. We then leverage the orthogonality of harmonic functions and obtain the following expression, valid for any integer $q \in \mathbb{Z}$:
\begin{equation}
    -\sum_{p=-\infty}^{\infty}\left( \omega+q\omega_m-p\omega_m \right)^2\hat{m}_{p}\hat{v}_{q-p}+\left(k+ic\left( \omega+q\omega_m \right)\right)\hat{v}_{q}\\
    =0.
    \label{e:exp}
\end{equation}

To solve Eq.~\ref{e:exp}, we have to truncate the summation by considering $|p| \le P$; similarly, we only consider $|q| \le Q$. Some algebra yields the following matrix equation:
\begin{equation}
    \left( \omega^2\mathbf{\hat{m}} + \omega\left(\omega_m \mathbf{Q_1} \circ \mathbf{\hat{m}}-ic\mathbf{I}\right) + \omega_m^2\mathbf{Q_2} \circ \mathbf{\hat{m}} - k\mathbf{I}-ic\omega_m\mathbf{Q_3}\right)\mathbf{\hat{v}}=\mathbf{0}_{(2Q+1)\times1},
    \label{e:eig2}
\end{equation}
where $\circ$ represents the Hadamard element-by-element product, $\mathbf{0}_{2Q+1\times1} \in \mathbb{R}^{2Q+1}$ is the zero vector and $\mathbf{I}\in \mathbb{R}^{(2Q+1)\,\times\,(2Q+1)}$ is the identity matrix. We define the vector of Fourier coefficients of the displacement amplitude, $\mathbf{\hat{v}} \in \mathbb{C}^{2Q+1}$, as
\begin{equation}
    \mathbf{\hat{v}}=\left[
    \begin{array}{ccccccc}
    \hat{v}_{-Q} & \hdots & \hat{v}_{-1} & \hat{v}_{0} & \hat{v}_{+1} & \hdots & \hat{v}_{+Q} 
    \end{array}
    \right]^\mathrm{T},
\end{equation}
with ``T'' being the matrix transposition.  \textcolor{black}{Matrix $\mathbf{\hat{m}}\in \mathbb{R}^{(2Q+1)\,\times\,(2Q+1)}$, that contains the Fourier coefficients of the effective mass, is given in ~\ref{a:mat}, together with the three matrices of coefficients $\mathbf{Q_1},\,\mathbf{Q_2},\,\mathbf{Q_3}\in \mathbb{R}^{(2Q+1)\,\times\,(2Q+1)}$.}

The eigenvalue problem in Eq.~\ref{e:eig2} is quadratic with respect to the eigenvalues $\omega$. We solve it numerically in MATLAB using the \verb+polyeig+ command, thus obtaining a set of eigenvalues and the corresponding eigenvectors.

\subsubsection{Response without forcing terms: Results}
\label{s:freeres}

To validate our analysis so far, we consider a structure with the following parameters: $k=1\,\mathrm{Nm^{-1}}$, $m=0.5\,\mathrm{kg}$, $m_i=m/2$, $l=1\,\mathrm{m}$, $\alpha_0=\pi/6$. Additionally, we choose a damping ratio $\xi=0.005$. In our problem, the damping ratio is defined as  \textcolor{black}{$\xi=c/(2 m_{e0} \omega_u)$}, where the characteristic frequency of the unmodulated system is  \textcolor{black}{$\omega_u=\sqrt{k/m_{e0}}$}, with  \textcolor{black}{$m_{e0}=m+m_i/(2\sin^2{\alpha_0})$ being the effective mass of Eq.~\ref{e:seom2} with modulation-related terms set to 0}. Regarding the modulation parameters, we set: $\omega_m=\omega_u/10$, $dv_b=0.1\,\mathrm{m}$. Keep in mind that $\omega_m$ needs to be much smaller than the characteristic frequency to avoid instabilities~\cite{Kovacic2018, Palermo2020}.  \textcolor{black}{For the interested reader, the time evolution of the time-modulated effective properties $m_{e}$ and $F_e$ for the chosen parameters is shown in ~\ref{a:eff}}.

\begin{figure}[!htb]
\centering
\includegraphics[scale=1.00]{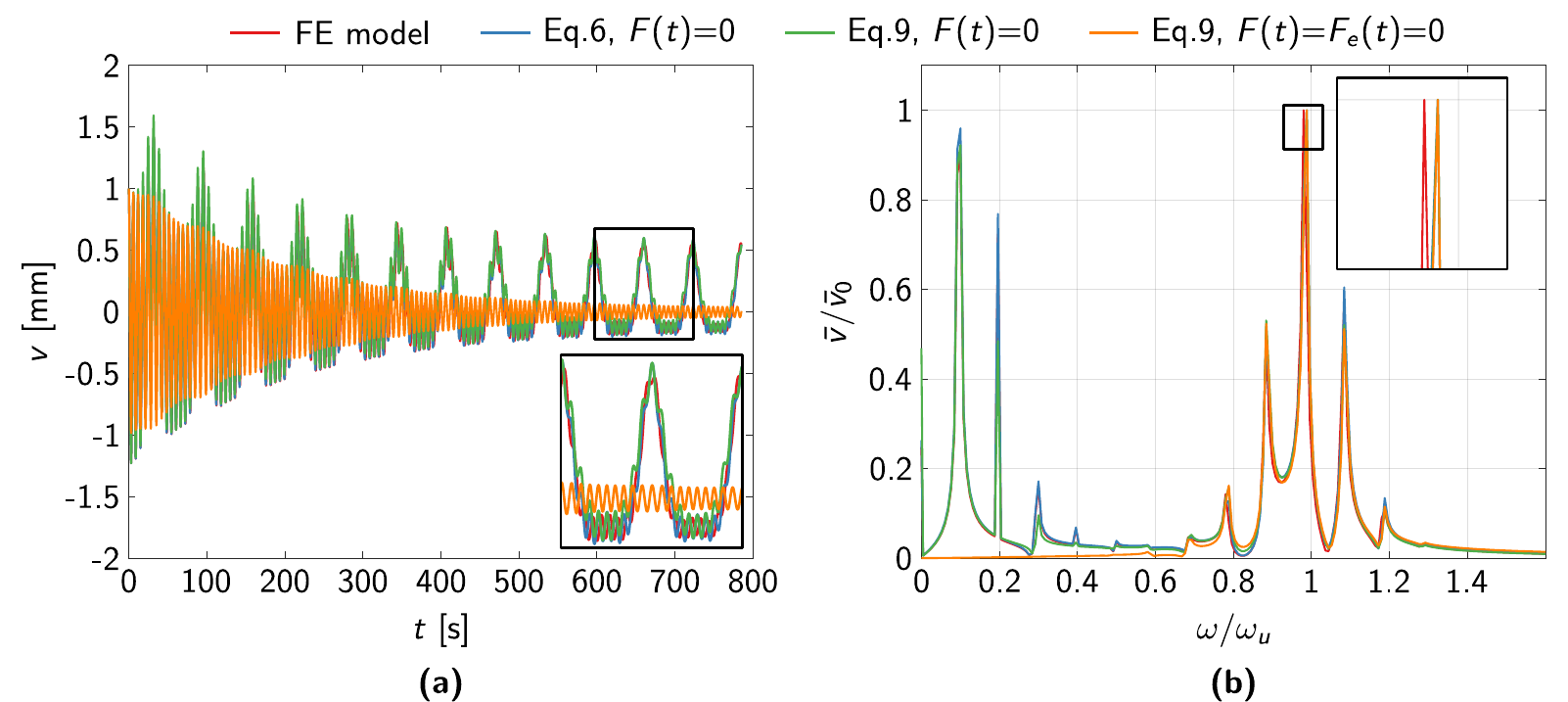}
\caption{Response of the time-modulated building block to initial conditions. (a) Time evolution. (b) Normalized frequency response, where $\bar{v}(\omega)$ is obtained from the simulated time histories via discrete Fourier transform. In each figure, we superimpose the response of the FE model (red line) to the numerical responses obtained by solving: the fully-nonlinear Eq.~\ref{e:nleom2} with $F(t)=0$ (blue line); the simplified Eq.~\ref{e:seom2} with $F(t)=0$ (green line), or with $F(t)=F_e(t)=0$ (orange line).}
\label{f:free1}
\end{figure}

As a first step, we illustrate the frequency response of our building block by means of numerical simulations, and provide detail on the influence of the various terms in our equations. In accordance with the assumption  $v \ll v_b$ of Section~\ref{s:simp}, we choose initial conditions $v(0)=0.001\,\mathrm{m}$ and $\dot{v}(0)=0\,\mathrm{ms^{-1}}$, such that the generated displacements are much smaller than the base displacement $dv_b$. We start from our most generic nonlinear equation, Eq.~\ref{e:nleom2}, and set $F(t)=0$. The time evolution and frequency content of the response obtained from this equation are shown as a blue line in Fig.~\ref{f:free1}(a) and (b), respectively. The equation is solved in Matlab with the \verb+ode45+ function, considering a time step size $dt=2\pi/(40\omega_u)$ and 5001 total time steps.
We observe that the frequency response has multiple peaks: a central one, accompanied by sidebands that are typical of modulated oscillators and that are approximately located at distances from the central peak that are multiple of the modulating frequency~\cite{Minkov2017, Palermo2020}. The central frequency, corresponding to the maximum amplitude of the response, is slightly smaller than the characteristic frequency $\omega_u$. Note that this deviation is inherent to the system and is not solely determined by damping. We also observe peaks at low frequency, that are approximately located at multiples of the modulating frequency. This result is compared to that obtained from a fully-nonlinear truss model implemented in the finite element (FE) software COMSOL Multiphysics (red line);  \textcolor{black}{see Section~\ref{s:FE results} for additional details on the COMSOL model}. We can see that the two curves overlap very well for low $t$, while some small discrepancies appear as the simulation goes on, as shown in the detail of Fig.~\ref{f:free1}(a). We can observe that their repercussions on the frequency content are minimal, with the most significant one represented by a slight shift of the main peak, as shown in the detail of Fig.~\ref{f:free1}(b). In the same figure, we also superimpose the response obtained by solving the simplified equation of motion, Eq.~\ref{e:seom2}, with $F(t)=0$ (green line). Again, we can see that the difference with the FE and the fully-nonlinear solution are minimal; this highlights that our simplifying assumptions are reasonable for our choice of parameters. 

Finally, we plot the response of the simplified equation of motion, Eq.~\ref{e:seom2}, with both $F(t)=0$ and the effective force given by the moving base set to zero: $F_e(t)=0$. In this case, the effect of the base modulation is entirely modeled as a time-varying mass. The corresponding time evolution (orange line) is notably different from the others. However, the frequency content highlights that this discrepancy mainly affects the low frequency behavior. Neglecting $F_e(t)$ causes the low-frequency peaks to disappear entirely, while it has little effects on the response around the characteristic frequency $\omega_u$. The consequences of the latter remark are significant. If we are interested in exciting the system at frequencies that are close to its characteristic frequency, as it is typically done in non-reciprocal devices, one can indeed model the effect of the inerter as a time-varying mass.

Given these results, we are ready to compare our numerical simulations to the eigenvalues obtained by solving Eq.~\ref{e:eig2}. It is particularly of interest to us to understand the role played by the truncation orders $P$ and $Q$. To avoid testing too many cases, we set $P=Q$. The eigenvalues for various values of $P=Q$ are compared to the peaks of the response obtained from our fully nonlinear equation of motion, Eq.~\ref{e:nleom2}, with $F(t)=0$.
\begin{table}
\small
\begin{tabular}{
  P{\dimexpr.16\textwidth-2\tabcolsep-1.3333\arrayrulewidth}
  !{\vrule width 1.5pt}P{\dimexpr.12\textwidth-2\tabcolsep-1.3333\arrayrulewidth}
  |P{\dimexpr.12\textwidth-2\tabcolsep-1.3333\arrayrulewidth}
  |P{\dimexpr.12\textwidth-2\tabcolsep-1.3333\arrayrulewidth}
  |P{\dimexpr.12\textwidth-2\tabcolsep-1.3333\arrayrulewidth}
  |P{\dimexpr.12\textwidth-2\tabcolsep-1.3333\arrayrulewidth}
  |P{\dimexpr.12\textwidth-2\tabcolsep-1.3333\arrayrulewidth}
  |P{\dimexpr.12\textwidth-2\tabcolsep-1.3333\arrayrulewidth}
  }
   & Sideb.\ -3 & Sideb.\ -2 & Sideb.\ -1 & Central & Sideb.\ +1 & Sideb.\ +2 & Sideb.\ +3 \\
 \ChangeRT{1.5pt}
 Eq.~\ref{e:nleom2}, $F(t)$=$0$ & 0.692  & 0.788  & 0.884  & 0.988  & 1.084 & 1.188  & 1.292   \\
 \hline
 Eq.~\ref{e:eig2}, $P$=$Q$=$1$ & \textcolor{gray}{--} & \textcolor{gray}{--} & 0.867  & 0.987  & 1.106  & \textcolor{gray}{--} & \textcolor{gray}{--}  \\
 \hline
 Eq.~\ref{e:eig2}, $P$=$Q$=$2$ & \textcolor{gray}{--} & 0.766 & 0.888 & 0.989 & 1.091 & 1.205  & \textcolor{gray}{--}  \\
 \hline
 Eq.~\ref{e:eig2}, $P$=$Q$=$3$ & 0.666  & 0.788  & 0.889  & 0.989 & 1.089 & 1.191  & 1.385  \\
\end{tabular}
 \caption{Normalized frequencies ($\omega/\omega_u$) of central peak and sidebands of the response of our modulated unit to initial conditions. We first display the peak locations extracted from the fully nonlinear Eq.~\ref{e:nleom2} with $F(t)=0$. These are compared to the eigenvalues obtained from Eq.~\ref{e:eig2}, by setting the truncation orders $P$ and $Q$ to different values. For convenience, we always choose $P=Q$.}
 \label{t:free}
\end{table}
First, we notice that the number of eigenvalues is related to the truncation order; a higher truncation order implies that more sidebands can be captured by our model. While the location of the analytically-determined central peak is similar to that of the nonlinear model for all orders of truncation, the quality of the sideband estimation increases by incresing the truncation order. Yet, even when choosing $P=Q=1$, the lowest truncation order, we see that the discrepancy in terms of the location of sidebands $\pm 1$ with respect to the fully nonlinear model is a reasonable 2\%. In light of this, we choose this truncation order throughout the rest of the article. 



\subsubsection{Forced response: Analytical derivations}
\label{s:forced}

We now proceed to solve Eq.~\ref{e:seoms} in the presence of forcing terms. Many of the steps in the following derivation are similar to those in a scenario without external forcing terms (Section~\ref{s:free}). We are particularly interested in studying the steady-state dynamics of this system. Thus, we assume the forcing term to be sinusoidal, $F(t)=F_0\,e^{i\Omega t}$, and the solution to be a harmonic at a frequency corresponding to that of the excitation signal, $\Omega$, and modulated by $\omega_m$:
\begin{equation}
v(t)=\left(\sum_{n=-\infty}^{\infty}\hat{v}_{n}\,e^{i n \omega_m t}\right)e^{i \Omega t},
\label{e:vf}
\end{equation}
with $\hat{v}_{n}$ being once again the Fourier coefficients of the displacement function.

Differentiating $v(t)$, plugging it into Eq.~\ref{e:seoms} together with Eq.~\ref{e:me} and Eq.~\ref{e:fe}, and simplifying $e^{i\Omega t}$, yields
\begin{equation}
    -\sum_{n=-\infty}^{\infty}\sum_{p=-\infty}^{\infty}\left( \Omega+n\,\omega_m \right)^2\hat{m}_{p}\hat{v}_{n}\,e^{i (n+p) \omega_m t}+\sum_{n=-\infty}^{\infty}\left(k+ic\left( \omega+n\omega_m \right)\right)\,\hat{v}_{n}\,e^{i n \omega_m t}=F_0+\sum_{p=-\infty}^{\infty}\hat{F}_{ep}e^{i p \omega_m t}e^{-i\Omega t}.
    \label{e:seriesf}
\end{equation}
Following the same operations we performed on Eq.~\ref{e:series}, we obtain the expression, valid for any integer $q \in \mathbb{Z}$,
\begin{multline}
    -\sum_{p=-\infty}^{\infty}\left( \Omega+q\omega_m-p\omega_m \right)^2\hat{m}_{p}\hat{v}_{q-p}+\left(k+ic\left( \omega+q\omega_m \right)\right)\hat{v}_{q}\\
    =F_0\left[ q=0 \right]+\sum_{p=-\infty}^{\infty}\hat{F}_{ep}\left( 1 \left[ p=q-\Omega/\omega_m \right] + \frac{\sin{\left( \pi \left(\Omega/\omega_m+p-q \right) \right)}}{\pi\left( \Omega/\omega_m+p-q \right)} \left[ p \neq q-\Omega/\omega_m \right] \right).
    \label{e:expf}
\end{multline}
In this expression, and in the following, we use Iverson brackets to compactly represent conditional statements; in other words $[S]=1$ if $S$ is true and $[S]=0$ if $S$ is false. The last term in Eq.~\ref{e:expf} accounts for the effective force given by the modulated mechanism. This conditional statement, that arises since this term is not a function of the Fourier integer of the solution ($n$) states that the contributions of this term are mainly felt when the excitation frequency $\Omega$ is an integer multiple of the modulation frequency $\omega_m$.

Truncating the summations such that $|p| \le P$ and $|q| \le Q$, we obtain the matrix equation
\begin{equation}
    \left( \Omega^2\mathbf{\hat{m}} + \Omega\left(\omega_m \mathbf{Q_1} \circ \mathbf{\hat{m}}-ic\mathbf{I}\right) + \omega_m^2\mathbf{Q_2} \circ \mathbf{\hat{m}} - k\mathbf{I}-ic\omega_m\mathbf{Q_3}\right)\mathbf{\hat{v}}=\mathbf{F}+\mathbf{\hat{F}},
    \label{e:forced}
\end{equation}
where the only difference from the free vibrations case (Eq.~\ref{e:eig2}) is represented by the presence of force vectors. Vector $\mathbf{F} \in \mathbb{R}^{2Q+1}$ contains the amplitude of the applied force and vector $\mathbf{\hat{F}} \in \mathbb{C}^{2Q+1}$ contains terms corresponding to the modulation-based effective force.  \textcolor{black}{These vectors are given in ~\ref{a:mat}.}

Eq.~\ref{e:forced} can be solved to extract the Fourier coefficients $\mathbf{\hat{v}}$ for any value of the excitation frequency $\Omega$. Each entry in $\mathbf{\hat{v}}$ represents the magnitude of a peak in the frequency response of the forced system, and corresponds to a specific frequency located at distances from the excitation frequency $\Omega$ that are integer multiples of the modulating frequency $\omega_m$. For example, $\hat{v}_{-Q}$ is located at $\Omega-Q\omega_m$.

\subsubsection{Forced response: Results}
\label{s:forced}

\begin{figure}[!htb]
\centering
\includegraphics[scale=1.00]{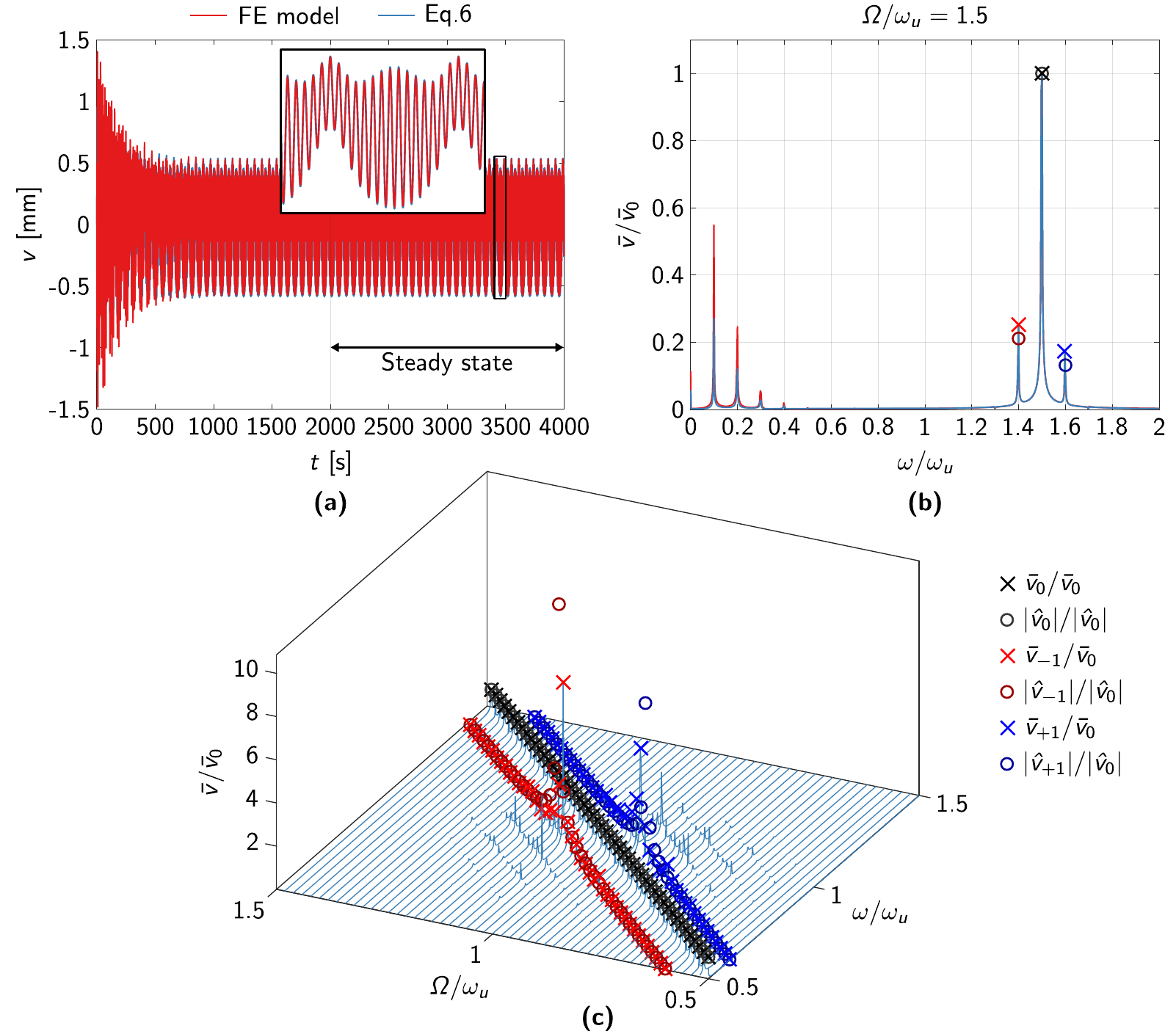}
\caption{Response of the time-modulated building block to a harmonic force applied at the main mass. (a) Time evolution of the response, with $\Omega=1.5\,\omega_u$. (b) Frequency content of the steady-state part of the response (a), i.e., the second half of the time history. The cross markers indicate the peaks of the response obtained by solving Eq.~\ref{e:nleom2} (blue line). The circular markers are the Fourier coefficients obtained by solving Eq.~\ref{e:forced}. Subscript 0 indicates the central peak, while -1 and +1 indicate the first left and right sidebands, respectively. The response is normalized by the amplitude of the peak at $\Omega$. (c) Evolution of the frequency response with $\Omega$.}
\label{f:forced}
\end{figure}

To validate our analysis, we consider the forced response of a system featuring the same parameters listed in Section~\ref{s:freeres}, with the addition of a small forcing amplitude $F_0=0.001\,\mathrm{N}$. Initially, we set $\Omega=1.5\,\omega_u$. First, we solve Eq.~\ref{e:nleom2} numerically using the same integration scheme and time discretization used in previous simulations. We obtain the time history shown as a blue line in Fig.~\ref{f:forced}(a).
This is compared to the time history obtained from a fully-nonlinear FE truss model (red line). We can see that the agreement between the two sets of results is excellent. From these time histories, we extract the steady-state part of the response (the second half of the time history), and obtain the frequency spectrum shown in Fig.~\ref{f:forced}(b); this is normalized by $\bar{v}_0$, the amplitude of the peak at the excitation frequency $\Omega$. In this spectrum, we can clearly see that the majority of the response is clustered around $\Omega$: there is a central peak at $\Omega$, and side peaks at $\Omega \pm \omega_m$. In addition, we observe some low-frequency peaks. We can see that the main discrepancies between our numerics and the FE model takes place away from the excitation frequency. The peaks from the numerical response of Eq.~\ref{e:nleom2} are indicated as cross-like markers for visualization purposes. 

We then solve Eq.~\ref{e:forced}, by setting the truncation order to $P=Q=1$; the Fourier coefficients we obtain are normalized by the Fourier coefficient of the central peak, $\hat{v}_0$. These coefficients are plotted at their respective frequencies in Fig.~\ref{f:forced}(b) and indicated by circular markers. We can see that the agreement between the Fourier coefficients and the numerical peaks is good, indicating the validity of our analytical procedure. In Fig.~\ref{f:forced}(c), we extend the comparison between numerical response and Fourier coefficients to other values of $\Omega$. We can see that the agreement between numerical peaks (crosses) and Fourier coefficients (circles) is excellent, with the largest deviations recorded when $\Omega$ is similar in value to the characteristic frequency of the system, $\omega_u$. 

\subsection{Remarks on feasibility and parameter selection}
\label{s:sdofremarks}

With the intent of providing information that could aid the design of non-reciprocal devices based on time-modulated inerters, we now study how some of the design parameters influence the response of our building block. Throughout this parametric study, unless otherwise specified, we adopt the same parameters listed in Section~\ref{s:freeres}. We limit ourselves to comparing frequency responses obtained by solving numerically the fully-nonlinear equation of motion (Eq.~\ref{e:nleom2}) with $F(t)=0$, and concentrate on a frequency interval in the neighborhood of $\omega_u$.
\begin{figure}[!htb]
\centering
\includegraphics[scale=1.00]{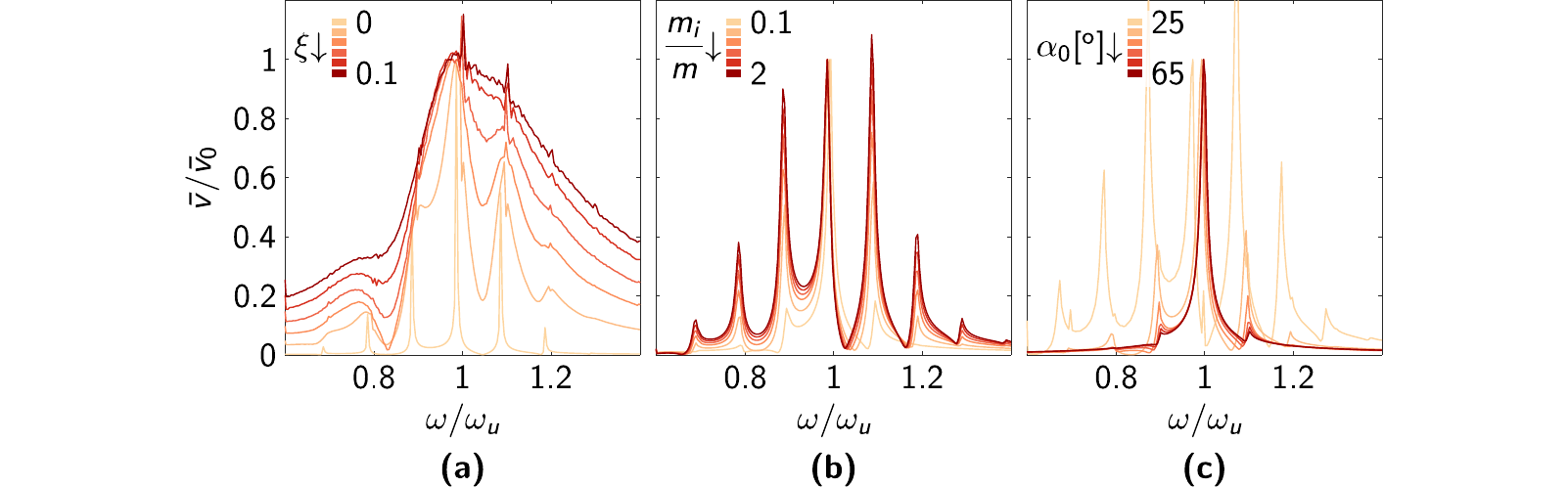}
\caption{Influence of the design parameters on the response of the time-modulated building block: (a) damping ratio, (b) mass ratio and (c) initial angle. }
\label{f:param}
\end{figure}
In the following parametric study, we postulate that the presence of prominent side peaks is an indicator of the strength of the time-modulation on the building block response. This information will be later used to guide the design a waveguide where  the time-modulated building block provides the desired non-reciprocal attributes.

We begin by studying the effects of the damping ratio $\xi$, by letting this parameter vary between 0 and 0.1, as shown in Fig.~\ref{f:param}(a). Here, we plot the Fourier amplitude normalized by its value at the peak closest to $\omega_u$ against the normalized frequency. As damping increases, the prominence of the side peak decreases to the point that, as we reach $\xi=0.1$, it is very hard to discern any side peak at all. We then investigate the effects of the mass ratio $m_i/m$, which we let vary between 0.1 and 2. As we increase this ratio, as seen in Fig.~\ref{f:param}(b), the side peaks increase in magnitude, highlighting that large lateral masses could be beneficial to achieve strong non-reciprocal effects.  Very significant is also the influence of the initial angle of the mechanism-based inerter, $\alpha_0$. Increasing this angle from 25 to 65 degrees, as shown in in Fig.~\ref{f:param}(c), the magnitude of the side peaks decreases considerably. In particular, for very low values of $\alpha_0$, which correspond to a configuration where the linkages of the mechanism are almost vertical, the amplitude of the side peaks becomes very large since decreasing $\alpha_0$ also causes the modulated part of the mass to increase in magnitude. 

We conclude this Section on the analysis of our modulated building block by formulating some remarks:
\begin{itemize}

    \item Damping is strongly detrimental to the effects of the base modulation.  
    
    \item Large lateral masses increase the modulation effect. Nonetheless, we need to keep in mind that increasing these masses causes the modulated component of the mass to increase in magnitude, and this could lead to instabilities.
    \item The initial geometrical configuration,  $\alpha_0$, has a major impact on the time-modulated response of the block; smaller $\alpha_0$ amplifies the modulation effect.
    \item The system's response is dictated by a multitude of different terms. The assumption $v \ll v_b$  simplifies the analysis of such systems and makes their response easy to predict with analytical approaches.
    \item Under the same $v \ll v_b$ assumption, the base modulation-generated forcing term $F_e(t)$ can be neglected if we operate near the characteristic frequency of the system $\omega_u$ and if $\omega_n \ll \omega_u$. This is significant since it allows to model the system as an effective time-modulated mass.
    \item Overall, it is best to work with small modulation frequencies, for the reason mentioned above and to avoid instabilities.
\end{itemize}

\section{Assembling building blocks into non-reciprocal devices: A non-reciprocal flexural waveguide}
\label{s:ass}

With a clear understanding of the dynamics of the modulated building block, we can now focus our investigation on its use as a modulated resonant element attached to an elastic waveguide.
As discussed in the introduction, several recent works have shown how beams, plates, and 2D substrates equipped with space-time modulated resonant elements can be used to achieve non-reciprocal wave phenomena \cite{Nassar2017eml,nassar2020,Palermo2020}. 

\begin{figure}[!htb]
\centering
\includegraphics[scale=1.00]{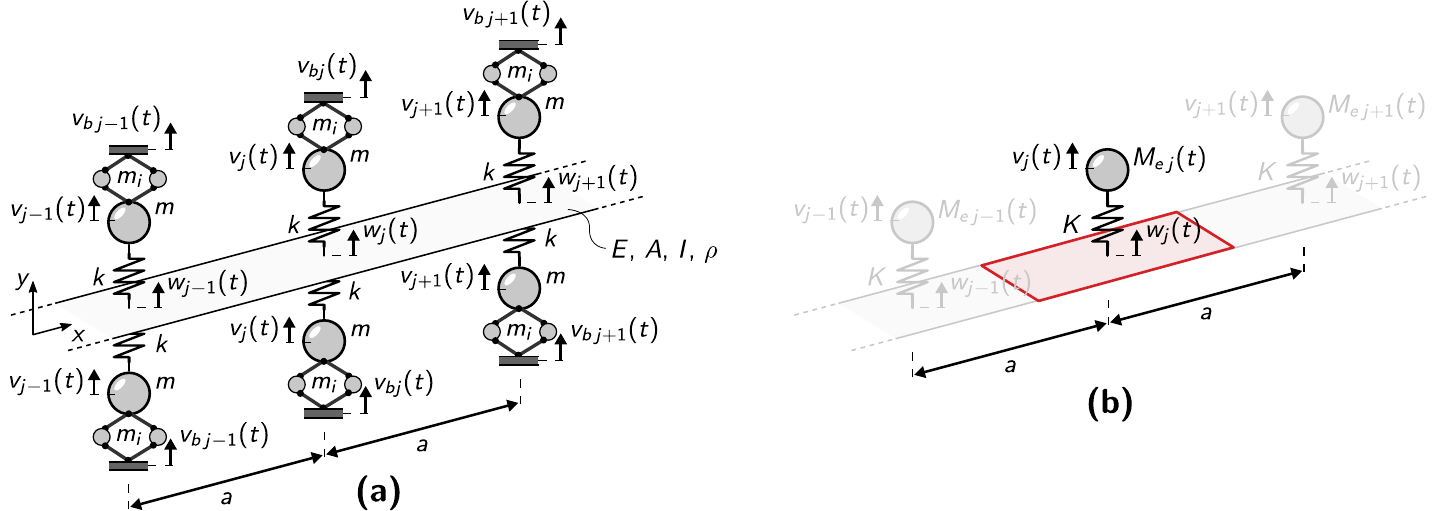}
\caption{(a) Portion of a flexural waveguide obtained by placing symmetric couples of identical building blocks at a constant distance $a$ along a beam. (b) Simplified model in which the base motion is approximated as a time-modulated mass; we highlight a single unit cell of this structure, while shading the rest in gray.}
\label{f:beam}
\end{figure}

 Here, we choose to demonstrate the capability of our structures to support the propagation of non-reciprocal flexural waves. To this purpose, we decorate an Euler Bernoulli beam with two arrays of our building blocks. The sketch of a portion of this flexural waveguide is shown in Fig.~\ref{f:beam}(a).
 The building blocks are attached to the two sides of the beam to form a symmetric distribution with respect to the beam axis (at $y=0$). Within both arrays, the base of each inerter can be actuated independently to obtain the desired spatio-temporal pattern across the beam. For all the examples below, we assume a right-traveling wave-like pattern for the inerter base motion, i.e.,
\begin{equation}
v_{b_j}(t)=\text{sign}(y)\, dv_b \cos(\omega_m t -\kappa_m x_j),
\label{e: in_vb}
\end{equation}
where $x_j$ is the coordinate of the $j$-th inerter and  $\kappa_m$ is the modulation wavenumber. Note that each block of the two symmetric arrays generates a force $F_e(x_j)$ which is equilibrated by its symmetric counterpart, thus avoiding the introduction of a net force on the flexural waveguide.

We remind the reader that the spatially-varying phase term, $\kappa_m x_j$,  introduces a periodic modulation that yields a symmetric bandgap $(BG(\kappa)=BG(-\kappa))$ in the dispersion relation of flexural waves. When combined with the temporal modulation, $\omega_m t$, these gaps are located at different frequencies for forward- and backward-traveling waves, thus giving rise to non-reciprocal effects \cite{Nassar2017eml}.

To investigate the expected non-reciprocal dispersive features, we resort to a simplified waveguide model, sketched in Fig.~\ref{f:beam}(b), where each couple of symmetric building blocks is substituted by a single oscillator with stiffness $K$, and effective time-varying mass $M_e$. The equivalent model allows us to predict the dispersive feature and the non-reciprocal behavior of the system using a standard truncated plane-wave expansion approach, as we shall see in Section~\ref{s:Dispersion}. In parallel, a finite portion of the waveguide equipped with the arrays of base-modulated building blocks is modeled  in a FE environment to provide evidence of the occurrence of non-reciprocal phenomena in the proposed flexural waveguide (see Section~\ref{s:FE results}).

\subsection{Dispersive properties}
\label{s:Dispersion}
We consider an Euler-Bernoulli beam with Young's Modulus $E$, cross-section area $ A $,  second moment of area $I$, and density $\rho$.
The beam is equipped with a linear array of oscillators with lattice distance $a$. As anticipated, each oscillator has an equivalent stiffness $K=2k$, and an effective mass $M_e(x_j,t)=2m_e(x_j,t)=2m_e(v_{bj}(x_j,t))$ dictated by the corresponding inerter base motion $v_{bj}(x_j,t)$ as per Eq.~\ref{e:seom}. In the long wavelength regime, the resonator effective mass $M_e(x_j,t)$ can be described as a function of the continuous variable $x$. Hence, expanding $M_e(x,t)$ in Fourier series and truncating its expression at the order $|p|=1$, we obtain:
\begin{equation}
M_e(t,x)\approx\sum_{p=-1}^{1}\hat{M}_{p}\,e^{ip( \omega_m t-\kappa_m x)}=\hat{M}_{0}+2d\hat{M}\cos(\omega_m t -\kappa_m x),
\label{e:mej}
\end{equation}
with $d\hat{M}=\hat{M}_{-1}=\hat{M}_{+1}$ and where $\hat{M}_{p}$ reads:
\begin{equation}
\hat{M}_{p}=\frac{\kappa_m}{2\pi} \frac{\omega_m}{2\pi} \int_{\frac{-\pi}{\kappa_m}}^{\frac{\pi}{\kappa_m}} \int_{\frac{-\pi}{\omega_m}}^{\frac{\pi}{\omega_m}}  M_e(x,t) \mathrm{e}^{-\mathrm{i}p (\omega_m t-\kappa_m x)}\,\mathrm{d}x\mathrm{d}t.
\end{equation}

\noindent For our numerical example, we employ the mechanical and modulation parameters of the building block described in Section~\ref{s:freeres} and set the mechanical parameters of the beam such that $\frac{\hat{M}_{0}}{\rho A a}=1.87$, with $\omega_m=0.1\omega_u$, $\frac{\kappa_u a}{2\pi}=0.098$, and $\kappa_m=1.86\kappa_u$, being $\kappa_u=\sqrt[4]{\frac{\omega_u^2\rho A }{EI}}$ the wavenumber corresponding to the pristine beam dispersion relation. 

We begin by briefly recalling the dispersive properties of a beam equipped with non-modulated resonators, e.g $M_e(t,x)=M$. Given the beam motion $w(x,t)$ and the array resonator motion $v(x,t)$, the governing equations of the flexural waveguide read
\begin{subequations}
\begin{align}
EI\frac{\partial^4 w(x,t)}{\partial x^4}+\rho A \frac{\partial^2 w(x,t)}{\partial t^2}+\frac{K}{a}(w(x,t)-v(x,t))=0,\\
M\frac{\partial^2 v(x,t)}{\partial t^2}-K(w(x,t)-v(x,t))=0.
\end{align}
\label{e:geb}
\end{subequations}
Note that the resonator mass motion at the discrete location $v_{j}(x_j,t)$ has been replaced by the continuous description $v(x,t)$, in accordance with the long wavelength assumption previously introduced.
At this stage, we assume a plane wave solution traveling along the flexural waveguide, such that
\begin{equation}
w(x,t)=w_0 \mathrm{e}^{\mathrm{i}(\omega t-\kappa x)}\quad\text{and}\quad
v(x,t)=v_0 \mathrm{e}^{\mathrm{i}(\omega t-\kappa x)},
\end{equation}
and the governing equations in Eq.~\ref{e:geb} thus read
\begin{subequations}
\begin{align}
\kappa^4 EI w_0-\omega^2\rho A w_0+\frac{K}{a}(w_0-v_0)=0,
-\omega^2 M v_0 - K(w_0-v_0)=0.
\end{align}
\end{subequations}
These can be rewritten in matrix form as
\begin{equation}
\boldsymbol{\Pi}(\kappa,\omega)\mathbf{V}=\mathbf{0},
\label{e:deb}
\end{equation}
with
\begin{equation}
\boldsymbol{\Pi}(\kappa,\omega)=\begin{bmatrix}
\kappa^4 EI  -\omega^2\rho A + K/a & -K/a\\
-K & -\omega^2 M + K\\
\end{bmatrix},\quad\mathbf{V}=\begin{bmatrix}
w_0\\
v_0\\
\end{bmatrix}
\label{e:gem}
\end{equation}
The dispersive solutions are thus found as the wavenumber-frequency ($\kappa,\omega$) pairs which satisfy $|\boldsymbol{\Pi}(\kappa,\omega)|=0$, namely the nontrivial solutions of Eq.~\ref{e:deb}. The reader can find all the details on the dispersive features of flexural waves along beams equipped with locally resonant elements in the literature~\cite{Yu2006}. Here, we simply recall that the dispersion curve  presents a gap around the resonator's natural frequency, which spans the band within $1<\omega/\omega_u<\sqrt{1+\beta}$, where $\beta=\frac{M}{\rho A a}$.

For the modulated waveguide, the plane wave solution is expanded to account for the scattered fields generated by the space-time modulation \cite{Nassar2017eml,Palermo2020}. The total field is thus obtained as
\begin{subequations}
\begin{align}
w(x,t)=\left(\sum_{n=-\infty}^{\infty}\hat{w}_{n}\,e^{i n( \omega_m t-\kappa_mx)}\right)e^{i( \omega t-\kappa x)}=\sum_{n=-\infty}^{\infty}\hat{w}_{n}\,e^{i n( \omega_n t-\kappa_nx)},\\
v(x,t)=\left(\sum_{n=-\infty}^{\infty}\hat{v}_{n}\,e^{i n( \omega_m t-\kappa_mx)}\right)e^{i( \omega t-\kappa x)}=\sum_{n=-\infty}^{\infty}\hat{v}_{n}\,e^{i n( \omega_n t-\kappa_nx)},
\end{align}
\label{e:edf}
\end{subequations}
where, for convenience, we have introduced the shifted frequency and wavenumber
\begin{equation}
\omega_n=\omega+n\omega_m\quad\text{and}  \quad \kappa_n=\kappa+n \kappa_m. 
\end{equation}
The governing equilibrium equations of the modulated flexural waveguide are thus obtained by substituting the expressions of the expanded displacement fields, Eq.~\ref{e:edf},  into the equilibrium equations, Eq.~\ref{e:geb}. In particular, by truncating the plane wave expansion such that $|n|\leq1$, we obtain the following matrix expression~\cite{Palermo2020}
\begin{equation}
\label{e:deb_mod}
\left[\begin{array}{ccc}
{\boldsymbol{\Pi}(\kappa_{-1},\omega_{-1})}&{\boldsymbol{\Gamma}(\kappa,\omega)} &\mathbf{0}\\
{\boldsymbol{\Gamma}(\kappa_{-1},\omega_ {-1})}&{\boldsymbol{\Pi}(\kappa,\omega)}&{\boldsymbol{\Gamma}(\kappa_{+1},\omega_{+1})}\\
\mathbf{0}&{\boldsymbol{\Gamma}(k,\omega)} &{\boldsymbol{\Pi}(\kappa_{+1},\omega_{+1})}
\end{array}\right]\left[\begin{array}{ccc}{\mathbf{\hat{V}}_{-1}}\\{\mathbf{\hat{V}}_{0}}\\ {\mathbf{\hat{V}}_{+1}}\end{array}\right]=\mathbf{0},
\end{equation}
where the submatrix $\boldsymbol{\Pi}(\kappa_n,\omega_n)$ has the same form as in Eq.~\ref{e:gem} with $M=\hat{M}_0$, while
\begin{equation} \label{e:Gamma}
\boldsymbol{\Gamma}(\kappa_n,\omega_n)=\left[\begin{array}{cc}
{0} & {0} \\
{0} & {-d\hat{M} \omega_n^2} \\
\end{array}\right],\quad\mathbf{\hat{V}_n}=\left[\begin{array}{cc}
\hat{w}_{n} \\
\hat{v}_{n}\\
\end{array}\right]
\end{equation}
From Eq.~\ref{e:deb_mod},  rewritten in compact form as $\mathbf{\Lambda}(\kappa,\omega)\,\mathbf{\hat{V}}=\mathbf{0}$, the dispersive solutions are obtained by setting the determinant  $|\mathbf{\Lambda}(\kappa,\omega)|=0$. 
As discussed in Ref.~\cite{Nassar2017eml}, the scattered fields, i.e., $\hat{w}_n$, $\hat{v}_n$ with $n\neq0$, are non-negligible only when the phase matching condition  $|\boldsymbol{\Pi}(\kappa,\omega)|$=$|\boldsymbol{\Pi}(\kappa_n,\omega_n)|$=0 is met \cite{Nassar2017prsa}, namely at the crossing points between the dispersion curve $|\boldsymbol{\Pi}(\kappa,\omega)|$=0 and the shifted twin curves $|\boldsymbol{\Pi}(\kappa,\omega)|$+$n(\kappa_m,\omega_m)$=0. Hence, the dispersive solutions, computed from $|\mathbf{\Lambda}(\kappa,\omega)|=0$ are valid only in the vicinity of such crossing points. 

For the sake of clarity, in the rest of the manuscript, we label with:
\begin{itemize}
    \item  $\mathcal{D}$ the dispersion curve of the modulated metabeam, obtained from the solutions of $|\mathbf{\Lambda}(\kappa,\omega)|=0$;
    \item $\mathcal{D}_0$ the dispersion curve of the fundamental branch, obtained from the solutions of $|\mathbf{\Pi}(\kappa,\omega)|=0$, with $M=\hat{M}_0$;
    \item $\mathcal{D}_{-1}$ the shifted dispersion curve obtained from the solutions of $|\mathbf{\Pi}(\kappa,\omega)|-(\kappa_m,\omega_m)=0$, with $M=\hat{M}_0$;
    \item $\mathcal{D}_{+1}$ the shifted dispersion curve  obtained from the solutions of $|\mathbf{\Pi}(\kappa,\omega)|+(\kappa_m,\omega_m)=0$, with $M=\hat{M}_0$;
\end{itemize}

\begin{figure}[!htb]
\centering
\includegraphics[scale=1.00]{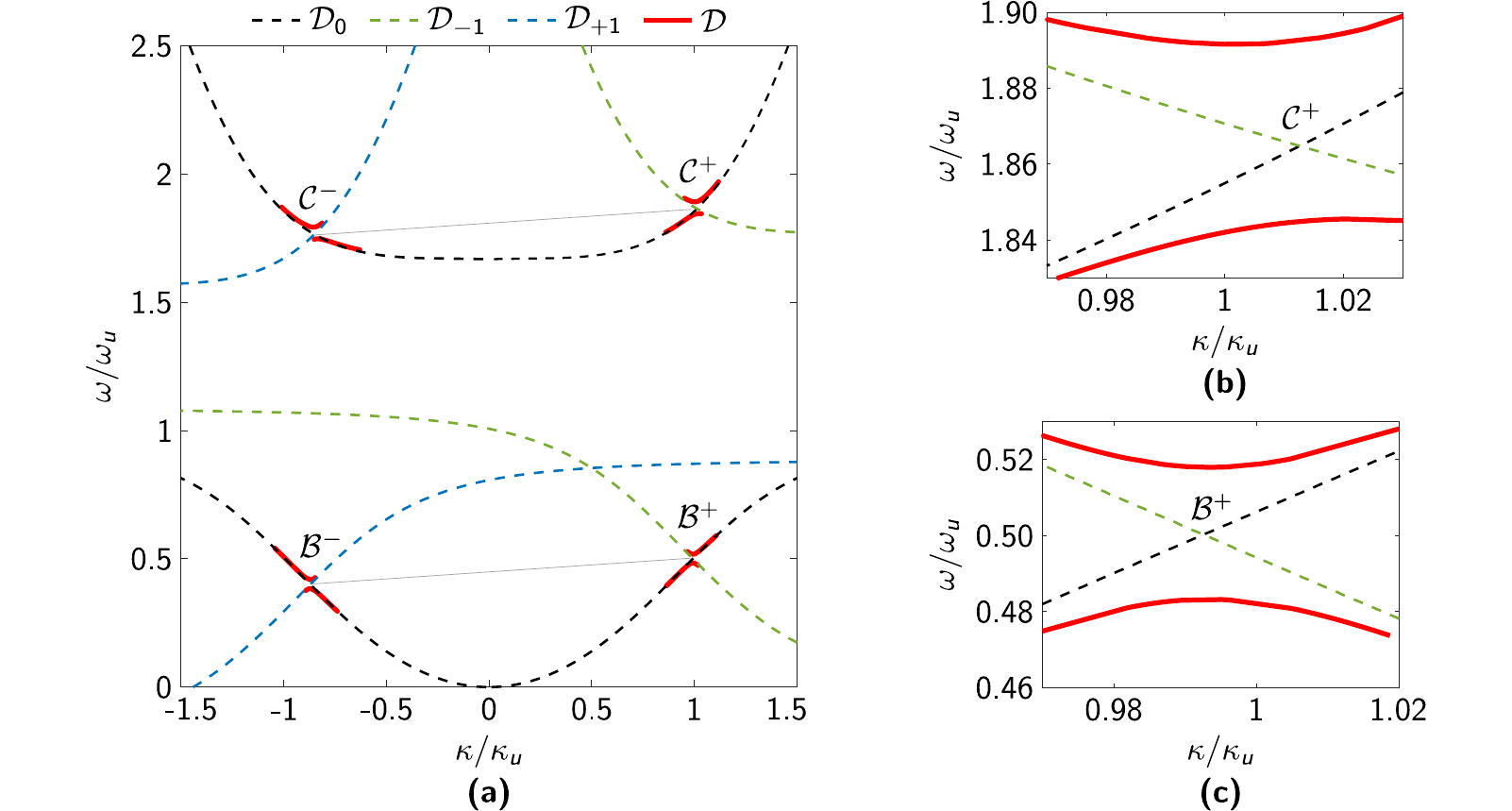}
\caption{Dispersive behavior of the modulated waveguide. (a) Dispersion curve of an Euler-Bernoulli beam with space-time mass modulated resonators. Red lines show the dispersion curve of the modulated waveguide, $\mathcal{D}$. Black lines mark the dispersive solutions of the fundamental modes $\mathcal{D}_0$, while green and blue lines mark the twin shifted curves $\mathcal{D}_{-1}$, $\mathcal{D}_{+1}$, respectively. (b), (c) Zoom-in of the partial bandgaps for right-going  ($\kappa>0$) flexural waves at $\omega/\omega_u=[0.48\text{--}0.52]$ and $\omega/\omega_u=[1.85\text{--}1.9]$. }
\label{f:disp}
\end{figure}

The curves $\mathcal{D}$, $\mathcal{D}_0$, $\mathcal{D}_{-1}$ and $\mathcal{D}_{+1}$, numerically computed for the adopted flexural waveguide, are reported in Fig.~\ref{f:disp}(a). 
In particular, the dispersion curve $\mathcal{D}$, computed in the vicinity of the phase matching point is shown with red lines.
The fundamental branch $\mathcal{D}_{0}$ and the shifted curves $\mathcal{D}_{-1}$ and $\mathcal{D}_{+1}$ are reported, instead, with black, green and blue lines, respectively.
As expected, the fundamental branches $\mathcal{D}_{0}$ and the shifted curves $\mathcal{D}_{-1}$, $\mathcal{D}_{+1}$ intersect at multiple locations where energy exchange between the fundamental branch and the scattered waves is expected. These locations can be collected in phase-matched pairs ~\cite{Nassar2017prsa} like $\mathcal{B}^{\pm}$ and $\mathcal{C}^{\pm}$ in Fig.~\ref{f:disp}, which are related by a translation $\pm(\kappa_m,\,\omega_m)$. A detailed discussion on the non-reciprocal wave propagation effects occurring across these points is provided in Ref.~\cite{Nassar2017eml}. Here, we simply recall that in the vicinity of one of these pairs, a directional band gap can be generated, as a result of the interaction between a fundamental ($\mathcal{D}_0$) and a scattered  ($\mathcal{D}_{\pm 1}$) mode having opposite group velocities. 
The reader can appreciate such directional gaps in the zoom-in of Fig.~\ref{f:disp}(b),(c).
Directional wave-filtering results from the reflection of the fundamental mode that is frequency-converted into the phase-matched mode~\cite{Nassar2017eml}.
For example, we expect a right-propagating wave with $\omega=1.87\omega_u$ to be reflected by the metabeam as a left-propagating wave with  $\omega=1.77\omega_u$. Numerical evidence of such non-reciprocal effects are provided in the next section.

\begin{figure}[!htb]
\centering
\includegraphics[scale=1.00]{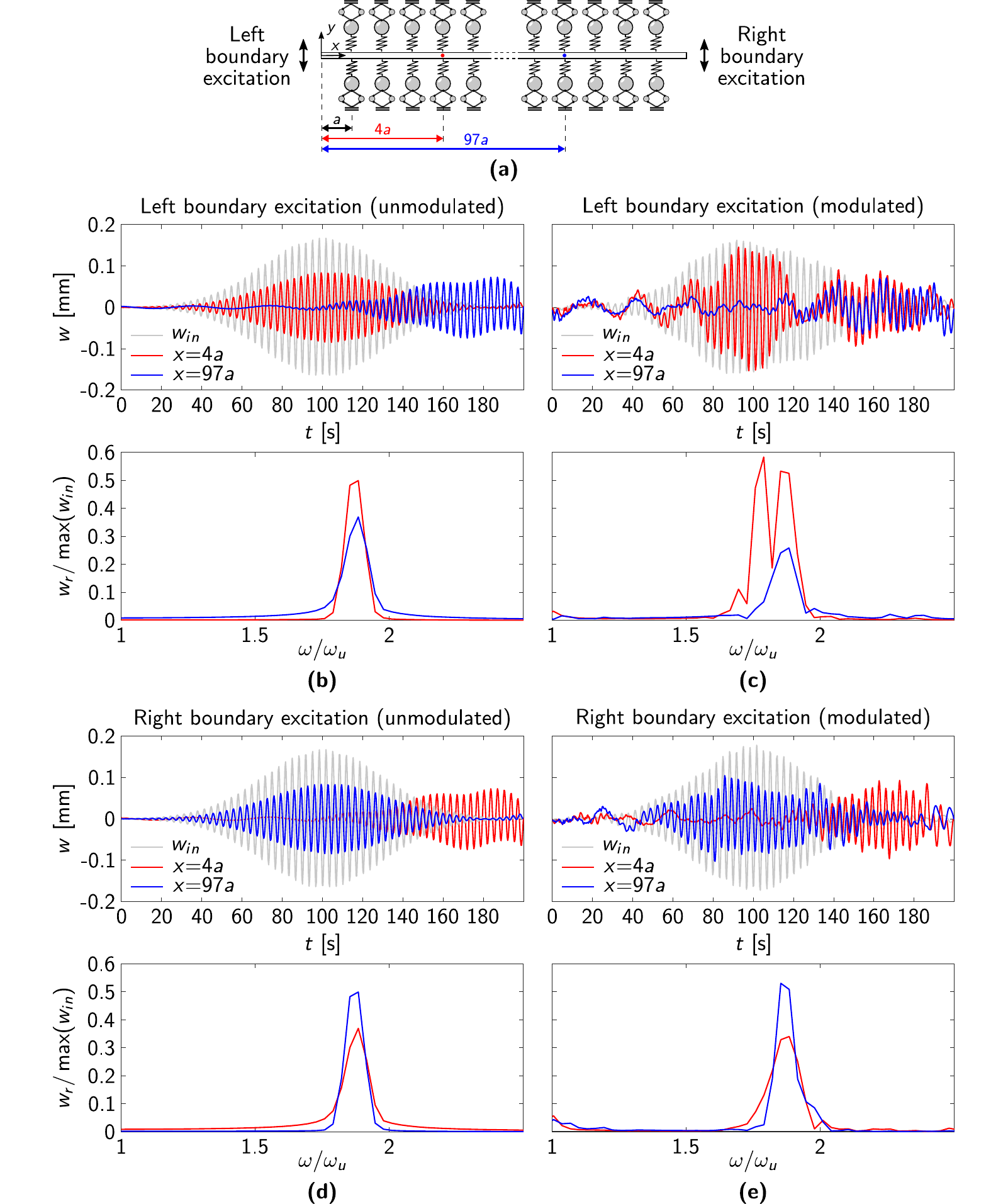}
\caption{ (a) Schematic of the simulated beam. Time histories and frequency spectra of the beam vertical displacements recorded at $x_r=4a$ and $x_r=97a$ for an excitation frequency $\Omega=1.87\omega_u$ and the following scenarios: right propagating waves in the (b) unmodulated and (c) modulated beam; left propagating waves in the (d) unmodulated and (e) modulated beam. In the time history plots, we also report the input signal $w_{in}$, recorded at the point of the beam closest to the excitation location.}
\label{f:transm}
\end{figure}

\subsection{Finite element modeling and results}
\label{s:FE results}
We model a finite portion of the Euler-Bernoulli beam equipped with symmetric arrays of inertially modulated resonators using a finite element approach. The modeled beam has an overall length $L=100a$ and hosts 99 pairs of modulated building blocks as shown in Fig.~\ref{f:transm}(a). The beam is modeled using standard beam elements (with cubic shape function) with a mesh length $s=a$ which ensures convergent results at the frequency of interest. Each building block includes 4 rigid connectors, modeled with truss elements of stiffness $K_c=500k$), a dashpot-spring element of stiffness $k$ and damping ratio $\xi=0.005$, and point masses for the resonator and the inerter. The base of each inerter is displaced as in Eq.~\ref{e: in_vb}, producing a right-going modulating wave.

We analyze the dynamic response of this finite waveguide using time-history analyses. To capture the nonlinear behavior of the base-displaced inerter, at each time step, the dynamic equilibrium equation is solved in the deformed configuration (as is the norm in geometrically nonlinear problems) using the Newton-Raphson algorithm.  \textcolor{black}{The adopted time marching algorithm is the generalized-alpha method. Note that the same numerical schemes are also used for all the FE simulations of the building block provided in Section~\ref{s:bb}}.

To obtain evidence of the expected non-reciprocal behavior, the beam is excited with a gaussian modulated harmonic signal $w_{in}(t)=W_{0}\sin(\Omega t)e^{-\frac{(t-t_0)^2}{2\sigma^2}}$, centered at the angular frequency $\Omega=1.87\omega_u$, i.e., within the partial-band gap for right propagating waves. In accordance  with the  assumption $v\ll v_b$, the input amplitude is set to $W_{0}=dv_b/500$, while the other parameters are defined as $t_0=\frac{50}{\pi}T_0$,  \textcolor{black}{with $T_0=\frac{2\pi}{\omega_u}$} and $\sigma=\frac{60\pi}{T_0}$. The excitation is applied at one of the beam edges keeping the other one fixed.

We begin our investigation by exciting the beam from the left boundary to obtain a right propagating wave at $\Omega=1.87\omega_u$, and by monitoring the response at receiving locations $x_r=4a$ and $x_r=97a$. Fig.~\ref{f:transm}(b)  \textcolor{black}{and (c) illustrate the time histories and frequency spectra of the recorded signals in the unmodulated and modulated case, respectively}. According to the dispersion curve in Fig.~\ref{f:disp}(a), a directional gap exists for right-propagating waves  \textcolor{black}{in the presence of a modulation}, and this generates back reflected waves at the paired frequency $\Omega-\omega_m=1.77\omega_u$. This is confirmed by looking at the frequency spectrum of the vertical beam displacement $w(x_r)$  recorded at the receiver $x_r=4a$  \textcolor{black}{(bottom panel of Fig.~\ref{f:transm}(c))}, which shows two distinct peaks associated with the incident wave at $1.87\omega_u$ and the back-reflected wave at $1.77\omega_u$. Conversely, the receiver at $x_r=97a$, located away from the source, shows a single peak at $1.87\omega_u$, related to the fundamental propagating mode. Note that the amplitude of this peak is significantly reduced with respect  \textcolor{black}{ to the same signal recorded in the unmodulated case (bottom panel of Fig.~\ref{f:transm}(b)),} as a result of the frequency conversion between the fundamental mode and the paired one.

To prove the directional behavior of the modulated waveguide, we excite the system from its right boundary and compute the frequency spectrum of the beam vertical displacement $w(x_r)$  recorded at the same receivers, as shown in Fig.~\ref{f:transm} \textcolor{black}{(d) and (e) for the unmodulated and modulated case, respectively}. In this scenario \textcolor{black}{, and in the presence of modulation}, the generated left-propagating waves can freely travel along the waveguide due to the absence of  \textcolor{black}{modulation-induced} gaps at $1.87\omega_u$, and no frequency-converted reflected waves are produced. This is confirmed by the spectra of the displacements at $x_r=4a$ and $x_r=97a$  \textcolor{black}{(bottom panel of Fig.~\ref{f:transm}(e))}, which only display components at the forcing frequency $\Omega=1.87\omega_u$. Moreover,  \textcolor{black}{unlike the right-propagating wave case}, one can appreciate that the amplitude away from the source is  \textcolor{black}{very close to that of the unmodulated counterpart (bottom panel of Fig.~\ref{f:transm}(d)), thus providing} further evidence of non-reciprocal dynamics.
 
\section{Conclusions}
\label{s:concl}

In this work, we have presented a time-varying mechanical device  realized with an inertially amplified system modulated by an imposed base displacement. By combining analytical investigations and numerical simulations, we have provided a detailed description of the system dynamics highlighting the main parameters driving its time-modulated response. In particular, we have shown that a careful selection of such parameters allows obtaining a dynamic response akin to an effective time-modulated mass resonator, ready to be employed for the design of waveguides with non reciprocal attributes. By means of wave expansion techniques and numerical simulations, we have also demonstrated the use of such time-modulated building block to realize a  modulated flexural waveguide which is able to support non-reciprocal  wave attenuation associated with frequency conversion.

 Overall, our proposed system presents two main advantages with respect to  other systems in the literature, which are inherited by the fact that the inerter element is connected to a moving base: 1) for some choices of parameters, the force term transmitted from the base to the main mass is negligible, thus avoiding small-on-large effects in our system, that can bury the propagating signal within a larger modulation signal; and  \textcolor{black}{2) by designing a camshaft that simultaneously actuates the bases of all building blocks in a metamaterial system, one could produce a spatiotemporal modulation and nonreciprocal effects with a single rotatory actuator. Thus, we believe that our findings will pave the way towards the realization of non-reciprocal waveguides that are closer to being fully-mechanical.}

\section*{Acknowledgments}
PC acknowledges the support of the Research Foundation for the State University of New York. AP acknowledges the support of DICAM at the University of Bologna. PC also thanks Myles Tucker for his support. 

\appendix
\setcounter{figure}{0} 

\section{Simplification of the nonlinear equation of motion}
\label{a:simp}

In this Appendix, we further explain our reasoning behind the assumptions that lead to the simplified equation of motion in Eq.~\ref{e:yeom}. In particular, our goal is to spell out a relationship between the order of terms $\mathrm{T}_3$, $\mathrm{T}_4$ and $\mathrm{T}_5$ of Eq.~\ref{e:nleom}. First of all, the assumption that $v \ll v_b$ can be re-written as: 
\begin{equation}
v_b=N\,v,
\label{e:N}
\end{equation}
where $N$ is large. In addition, the assumption that $\omega$, the system's operating frequency, is much larger than the modulation frequency $\omega_m$ can be rewritten as 
\begin{equation}
\omega=M\,\omega_m,
\label{e:M}
\end{equation}
where $M$ is also large. Assuming that $v_b$ is a harmonic function of $\omega_m$, and using Eq.~\ref{e:N}, we can write: 
\begin{equation}
    \dot{v}_b=\omega_m v_b=N\omega_m v.
\end{equation}
Similarly, assuming that the main mass of the system will vibrate at the operating frequency $\omega$, and using Eq.~\ref{e:M}, we can write:
\begin{equation}
    \dot{v}=\omega v=M\omega_m v.
\end{equation}
In light of this, we can conclude that $(\dot{v}-\dot{v}_b)^2$ the driving term of $\mathrm{T}_3$, $\mathrm{T}_4$, can be rewritten as:
\begin{equation}
    (\dot{v}-\dot{v}_b)^2=(M-N)^2\omega_m^2v^2.
\end{equation}
Since the second derivative of $v_b$ can be written as $\ddot{v}_b=\omega_m^2v_b=N\omega_m^2v$ and $\ddot{v}=\omega^2v=M^2\omega_m^2v$, the driving term of $\mathrm{T}_5$ can be rewritten as:
\begin{equation}
    \ddot{v}-\ddot{v}_b=(M^2-N)\omega_m^2v^2.
\end{equation}
Assuming $M$ and $N$ to be of the same order, we can conclude that $(M^2-N) \gg (M-N)^2$ and, thus, $|\ddot{v}-\ddot{v_b}| \gg |(\dot{v}-\dot{v_b})^2|$. This justifies neglecting $\mathrm{T}_3$ and $\mathrm{T}_4$ in Eq.~\ref{e:nleom}.

\section{ Matrices for Eq.~\ref{e:eig2} and Eq.~\ref{e:forced}}
\label{a:mat}

{ In this Appendix, we report the definitions of some of the matrices used in Eq.~\ref{e:eig2} and Eq.~\ref{e:forced} -- which are not reported in the main text to improve readability. Matrix $\mathbf{\hat{m}}\in \mathbb{R}^{(2Q+1)\,\times\,(2Q+1)}$ contains the Fourier coefficients of the effective mass, as follows:
\begin{equation}
    \mathbf{\hat{m}}=\left[
\begin{array}{ccccccc} 
    \hat{m}_{0} & \hdots & \hat{m}_{-Q+1} & \hat{m}_{-Q} & \hat{m}_{-Q-1} & \hdots & \hat{m}_{-2Q}\\
    \vdots & \ddots & \vdots & \vdots & \vdots &  & \vdots\\
    \hat{m}_{+Q-1} & \hdots & \hat{m}_{0} & \hat{m}_{-1} & \hat{m}_{-2} & \hdots & \hat{m}_{-Q-1}\\
    \hat{m}_{+Q} & \hdots & \hat{m}_{+1} & \hat{m}_{0} & \hat{m}_{-1} & \hdots & \hat{m}_{-Q}\\
    \hat{m}_{+Q+1} & \hdots & \hat{m}_{+2} & \hat{m}_{+1} & \hat{m}_{0} & \hdots & \hat{m}_{-Q+1}\\
    \vdots &  & \vdots & \vdots & \vdots & \ddots & \vdots\\
    \hat{m}_{+2Q} & \hdots & \hat{m}_{+Q+1} & \hat{m}_{+Q} & \hat{m}_{+Q-1} & \hdots & \hat{m}_{0}
\end{array}
\right].
\end{equation}
Note that some Fourier coefficients $\hat{m}_i$ are zero when $|i|>P$. The three matrices of coefficients $\mathbf{Q_1},\,\mathbf{Q_2},\,\mathbf{Q_3}\in \mathbb{R}^{(2Q+1)\,\times\,(2Q+1)}$ are defined as follows:
\begin{gather}
\mathbf{Q_1}=\left[
\begin{array}{ccccccc} 
    -2Q & \hdots & -2 & 0 & 2 & \hdots & 2Q\\
    \vdots & \ddots & \vdots & \vdots & \vdots &  & \vdots\\
    -2Q & \hdots & -2 & 0 & 2 & \hdots & 2Q\\
    -2Q & \hdots & -2 & 0 & 2 & \hdots & 2Q\\
    -2Q & \hdots & -2 & 0 & 2 & \hdots & 2Q\\
    \vdots &  & \vdots & \vdots & \vdots & \ddots & \vdots\\
    -2Q & \hdots & -2 & 0 & 2 & \hdots & 2Q\\
\end{array}
\right], \qquad 
\mathbf{Q_2}=\left[
\begin{array}{ccccccc} 
    (-Q)^2 & \hdots & (-1)^2 & 0 & 1^2 & \hdots & Q^2\\
    \vdots & \ddots & \vdots & \vdots & \vdots &  & \vdots\\
    (-Q)^2 & \hdots & (-1)^2 & 0 & 1^2 & \hdots & Q^2\\
    (-Q)^2 & \hdots & (-1)^2 & 0 & 1^2 & \hdots & Q^2\\
    (-Q)^2 & \hdots & (-1)^2 & 0 & 1^2 & \hdots & Q^2\\
    \vdots &  & \vdots & \vdots & \vdots & \ddots & \vdots\\
    (-Q)^2 & \hdots & (-1)^2 & 0 & 1^2 & \hdots & Q^2\\
\end{array}
\right]\nonumber\\
\mathbf{Q_3}=\left[
\begin{array}{ccccccc} 
    -Q & \hdots & 0 & 0 & 0 & \hdots & 0\\
    \vdots & \ddots & \vdots & \vdots & \vdots &  & \vdots\\
    0 & \hdots & -1 & 0 & 0 & \hdots & 0\\
    0 & \hdots & 0 & 0 & 0 & \hdots & 0\\
    0 & \hdots & 0 & 0 & 1 & \hdots & 0\\
    \vdots &  & \vdots & \vdots & \vdots & \ddots & \vdots\\
    0 & \hdots & 0 & 0 & 0 & \hdots & Q
\end{array}
\right].
\end{gather}

Finally, vectors $\mathbf{F} \in \mathbb{R}^{2Q+1}$ and vector $\mathbf{\hat{F}} \in \mathbb{C}^{2Q+1}$, that contain the amplitude of the applied force and terms corresponding to the modulation-based effective force, respectively, and that are both used in Eq.~\ref{e:forced}, are defined as
\begin{equation}
\mathbf{F}=\left[
\begin{array}{ccccccc} 
    0\\
    \vdots\\
    0\\
    F_0\\
    0\\
    \vdots\\
    0\\
\end{array}
\right], \qquad 
\mathbf{\hat{F}}=\sum_{p=-\infty}^{\infty}\hat{F}_{ep}\left[
\begin{array}{ccccccc} 
    1 \left[ p=-Q-\Omega/\omega_m \right] + \frac{\sin{\left( \pi \left(\Omega/\omega_m+p+Q \right) \right)}}{\pi\left( \Omega/\omega_m+p+Q \right)} \left[ p \neq -Q-\Omega/\omega_m \right]\\
    \vdots\\
    1 \left[ p=-1-\Omega/\omega_m \right] + \frac{\sin{\left( \pi \left(\Omega/\omega_m+p+1 \right) \right)}}{\pi\left( \Omega/\omega_m+p+1 \right)} \left[ p \neq -1-\Omega/\omega_m \right]\\
    1 \left[ p=-\Omega/\omega_m \right] + \frac{\sin{\left( \pi \left(\Omega/\omega_m+p \right) \right)}}{\pi\left( \Omega/\omega_m+p \right)} \left[ p \neq -\Omega/\omega_m \right]\\
    1 \left[ p=1-\Omega/\omega_m \right] + \frac{\sin{\left( \pi \left(\Omega/\omega_m+p-1 \right) \right)}}{\pi\left( \Omega/\omega_m+p-1 \right)} \left[ p \neq 1-\Omega/\omega_m \right]\\
    \vdots\\
    1 \left[ p=Q-\Omega/\omega_m \right] + \frac{\sin{\left( \pi \left(\Omega/\omega_m+p-Q \right) \right)}}{\pi\left( \Omega/\omega_m+p-Q \right)} \left[ p \neq Q-\Omega/\omega_m \right]\\
\end{array}
\right].
\end{equation}
}

{ 
\section{Plot of the effective parameters as a function of time}
\label{a:eff}

For completeness, in Fig.~\ref{f:eff}, we plot the effective mass $m_e$ and effective force $F_e$ for the choice of system parameters in Section~\ref{s:freeres}. We can see that the effective mass oscillates about the non-modulated value $m_{e0}$, while the effective force oscillates about 0.
\begin{figure}[!htb]
\centering
\includegraphics[scale=1.00]{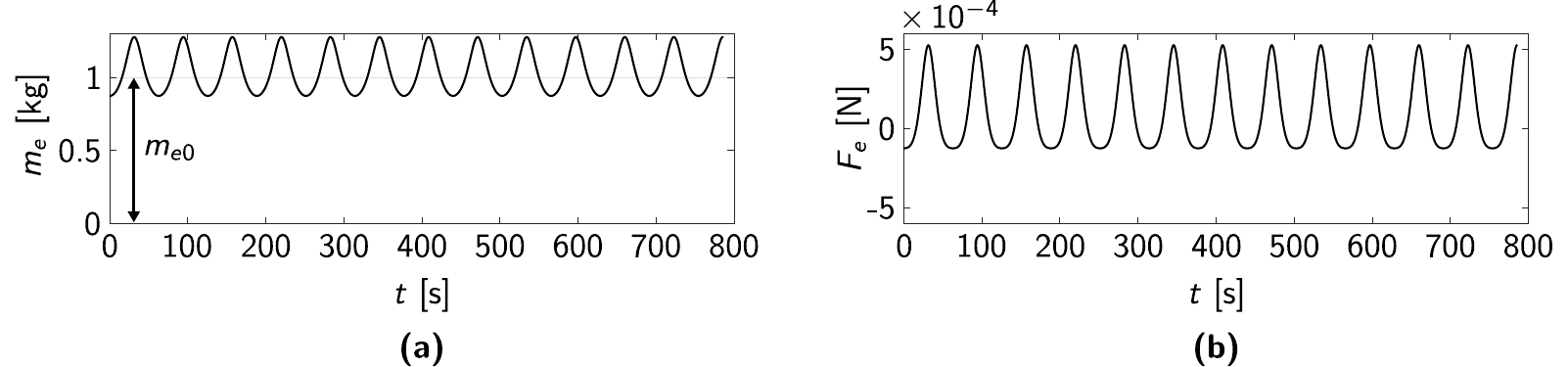}
\caption{ (a) Effective mass $m_e$ and effective force $F_e$ due to the base modulation, as a function of time, for the parameters given in Section~\ref{s:freeres}.}
\label{f:eff}
\end{figure}
}


\end{document}